\documentclass{aa} 

\usepackage{booktabs}
\usepackage{graphicx}
\usepackage{txfonts}
\usepackage{gensymb}
\usepackage{array}
\usepackage{hyperref}
\usepackage{graphicx}
\usepackage{color}

\usepackage{orcidlink}
\usepackage{multirow}

\begin{document} 

\title{First coordinated observations between Solar Orbiter\\ and the Daniel K. Inouye Solar Telescope}

\titlerunning{Solar Orbiter and DKIST}

\authorrunning{Barczynski, K., et al.}

 \author{Krzysztof Barczynski
 \inst{1,2}\orcidlink{0000-0001-7090-6180},
 Miho Janvier\inst{3}\orcidlink{0000-0002-6203-5239},
 Chris J. Nelson\inst{3}\orcidlink{0000-0003-1400-8356},
 T. Schad\inst{4}\orcidlink{0000-0002-7451-9804},
 A. Tritschler\inst{5}\orcidlink{0000-0003-3147-8026},
 Louise Harra\inst{2,1}\orcidlink{0000-0001-9457-6200},
 Daniel M\"uller\inst{3}\orcidlink{0000-0001-9027-9954},
 Susanna Parenti\inst{6} \orcidlink{0000-0003-1438-1310},
 Gherardo Valori\inst{7}\orcidlink{0000-0001-7809-0067},
 Gianna Cauzzi\inst{5}\orcidlink{0000-0002-6116-7301}
 and
 Yingjie Zhu\inst{1, 2}\orcidlink{0000-0003-3908-1330}
 }

\institute{ETH-Z\"urich, H\"onggerberg campus, HIT building, Wolfgang-Pauli-Str. 27, 8093 Z\"urich, Switzerland
\\
\email{krzysztof.barczynski@pmodwrc.ch}
\and
PMOD/WRC, Dorfstrasse 33, CH-7260 Davos Dorf, Switzerland
\and 
European Space Agency (ESA), European Space Research and Technology Centre (ESTEC), Keplerlaan 1, 2201 AZ Noordwijk, The Netherlands 
\and 
National Solar Observatory, 22 \'{}\={O}hi\'{}a K\={u} Street, Pukalani, HI 96768, USA
\and 
National Solar Observatory, 3665 Discovery Drive, Boulder, CO, 80303, USA
\and
Universit\'e Paris–Saclay, CNRS, Institut d'astrophysique spatiale, 91405, Orsay, France
\and 
Max Planck Institute for Solar System Research, Justus-von-Liebig-Weg 3, 37077 G\"ottingen, Germany
}
 
\date{Received xxx; accepted xxx}

\abstract
{Solar Orbiter and the Daniel K. Inouye Solar Telescope (DKIST) are two of the newest facilities available to the solar physics community. The first coordinated observations of the Sun by these two facilities occurred over the course of one week in October 2022. The returned data are open-access and will provide a valuable resource to researchers in the field.}
{Here, we provide an overview of the datasets collected by Solar Orbiter and DKIST through this coordination and discuss their scientific potential. Our aim is to demonstrate how these unique high-resolution coordinated observations, as well as similar observations obtained through subsequent campaigns, can help tackle important science questions in the field.}
{A decayed active region (without NOAA number during our observation but identified as AR13110 during previous solar rotation) was simultaneously observed by Solar Orbiter and DKIST at specific times between 18 and 24 October 2022. Between these dates, the Solar Orbiter spacecraft moved from a position with a separation angle of $77^\circ$ with Earth to a position with a separation angle of $51^\circ$, allowing stereoscopic observations to be collected with the ground-based telescope DKIST. From Solar Orbiter, observations are provided by the Extreme Ultraviolet Imager (EUI), Polarimetric and Helioseismic Imager (PHI), and the Spectral Imaging of the Coronal Environment (SPICE) instruments. Meanwhile, DKIST observed using the Cryogenic Near Infrared Spectropolarimeter (CryoNIRSP), the Visible Broadband Imager (VBI), and the Visible Spectropolarimeter (ViSP).}
{Coordinated observations were successfully collected at several distinct times over the week. Despite the active region itself being in an advanced decayed phase, a range of interesting features are evident in the collected data. As such, a variety of research topics can be advanced using these observations. In this article, we focus on three specific topics as representative examples, namely, coronal loop physics, the formation and evolution of the small-scale active region brightenings, and coronal rain dynamics.}
{The first coordinated observation campaign conducted by both Solar Orbiter and DKIST was a success. These open-access observations, and others like them, should help the solar physics community tackle key questions in the field. Such stereoscopic coordinated observations open up a new era in the analysis of the solar atmosphere.}


\maketitle

\section{Introduction}
\label{sect:intro}

The solar atmosphere is a complex environment that is both highly structured, due to the strong magnetic field that permeates throughout it, and stratified, with temperatures and densities that vary by several orders of magnitude as one moves away from the solar surface. Due to this complexity, obtaining a complete picture of the solar atmosphere from a single telescope is not a trivial task. Because of absorption effects in the Earth's atmosphere, ground-based assets are limited compared to space-based instruments. The atmospheric absorption restricts the spectral range of ground-based instruments, but they still have advantages, including the potential for larger apertures that allow for a higher spatial resolution and easier upgrades. In contrast, space-based instruments offer access to the complete electromagnetic spectrum; however, they are limited by different constraints, such as instrument weight, volume, and telemetry. Complementary observations from ground-based and space-based instruments therefore allow the strengths of both types of instruments to be used, mitigating their respective limitations and providing the most comprehensive observational solutions. As such, coordinating observations from multiple instruments (both on the ground and in space) is essential to understanding the transfer of mass and energy between different regions of the solar atmosphere. Here, we describe the first coordinated observations between two of the newest ground-based and space-borne infrastructure available to the solar physics community, namely, the U.S. National Science Foundation's Daniel K. Inouye Solar Telescope \citep[DKIST;][]{Rimmele2020} and the ESA/NASA Solar Orbiter \citep{Muller2020}.

\begin{figure}[!htb]
\center{\includegraphics[width=8.8cm]{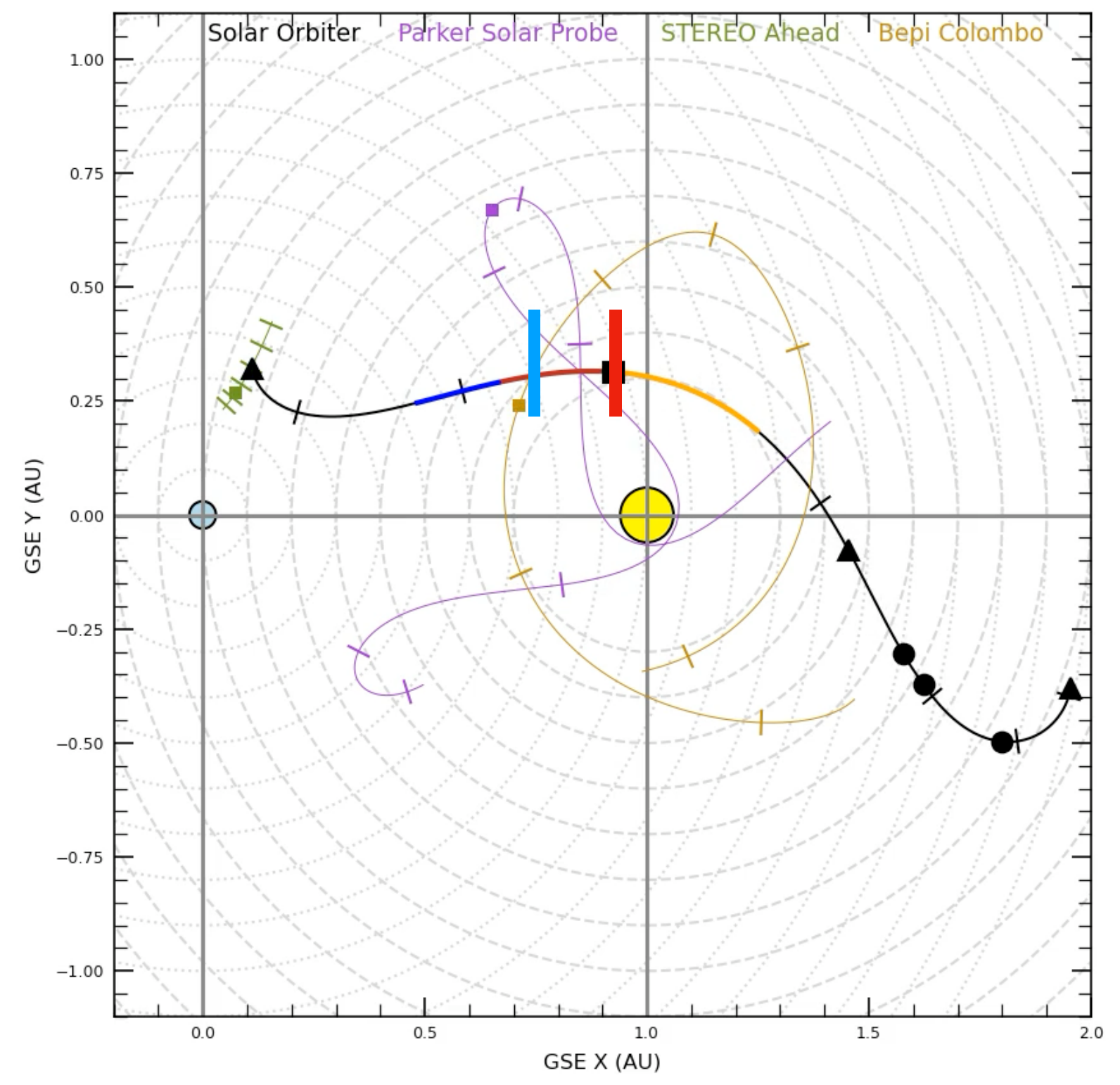}
\caption{Trajectory of Solar Orbiter (black line) with respect to the Sun and Earth in Geocentric Solar Ecliptic coordinates between 27 June 2022 (right-hand end of the line) and 26 December 2022 (left-hand end of the black line). The remote sensing windows (RSWs) of Solar Orbiter are indicated by the orange (RSW-4), red (RSW-5), and blue (RSW-6) parts of the trajectory. The RSWs started on 8 October and ended on 7 November, with the duration of each RSW being 10 days. The coordinated observations between DKIST and Solar Orbiter happened during RSW-5 and are indicated by the vertical lines of bold red (first observation from DKIST on 18 October 2022) and blue (last observation from DKIST on 24 October 2022). For reference, the plot also contains the trajectories of the ESA/JAXA mission Bepi Colombo (brown line), the NASA missions STEREO-A (green line), and the Parker Solar Probe (purple line) during the same half-year period.}
\label{fig:ar_overview_position_intro}} 
\end{figure}

The ESA/NASA Solar Orbiter spacecraft was launched in early 2020, and it started its nominal phase at the end of 2021. Solar Orbiter is a complex and ambitious mission with a unique trajectory that carries the spacecraft to a distance of only 0.28 au from the Sun at its closest approach. On board sits a suite of ten instruments (six remote sensing and four in situ), and each probes different regions of the Sun and the inner heliosphere. Although each instrument is capable of returning exciting data in isolation, it is often through combining them that we are able to truly investigate the biggest open questions in the field. In early 2022, the first Solar Orbiter Observing Plans (SOOPs; see \citealt{Zouganelis2020}) were executed, which coordinated observations from the different instruments on board the spacecraft to fulfill common science objectives. From 8 October until 7 November 2022, another set of remote-sensing observation windows (three windows of 10 days each) was carried out, collecting further data for several SOOPs. During that time, the spacecraft moved from slightly behind the Sun (from the Earth's perspective) to a position in front of the Sun, where co-observations with ground-based infrastructures were possible (see Fig.~\ref{fig:ar_overview_position_intro} for a schematic of the trajectory of Solar Orbiter during the six months around these dates). 

Upon completion of the construction phase in 2021 and its transition into the operations commissioning phase (OCP),  DKIST became the largest optical ground-based solar telescope in the world.
 DKIST is a four-meter diameter ground-based off-axis Gregorian telescope designed to conduct high-accuracy imaging and polarimetry at visible and infrared wavelengths. The telescope design allows for both on-disc observations (with the highest spatial resolution achieved to date by the solar physics community; \citet{Woeger2021}) and off-disc observations (both spectroscopic and spectropolarimetric; \citet{Rimmele2020}) measurements to be routinely collected. (See \citet{Rast2021} for a summary of the key science questions that will be addressed by DKIST.) 
Early science opportunities and community participation during the OCP are facilitated through the submission of PI-led proposals in response to OCP proposal calls. The Cycle 1 Proposal Call offered three of DKIST’s five first-generation instruments with limited capabilities. The respective Cycle 1 proposal-based observations started in early 2022. Although co-observing efforts are not part of the OCP (and as such not available in regular observing time), the unique science opportunities that these observations offer the community combined with the need to develop and test such activities were recognised in the development phases of both facilities. In order to enable such efforts and allow for faster data distribution to the international community, all co-observing efforts were pre-approved by DKIST Directorate and planned and executed outside of the proposal cycle framework. 

In this article, we introduce the data products collected by these two facilities during their first coordinated observations. The Solar Orbiter observations took place during an instance of the Long-Term Active Region SOOP,\footnote{\url{https://s2e2.cosmos.esa.int/confluence/display/SOSP/R_SMALL_MRES_MCAD_AR-Long-Term}} the broad aim of which is to resolve dynamic structures at small length and timescales in the solar atmosphere across a broad temperature range (from the photosphere to the corona) in a single active region over multiple days to understand the changes in an active region. While the Long-Term Active Region SOOP can target well-developed regions, with flaring events \citep[e.g.][]{Janvier2023}, the aim of the October 2022 SOOP run was to target the decaying phase.
The coordinated observations collected by DKIST and Solar Orbiter are open-source and will help the community study the evolution of a variety of phenomena that occur in active regions. Here, we discuss three representative research questions to demonstrate how  Solar Orbiter and  DKIST coordinated observations can advance our understanding of the Sun. These questions are as follows:
\begin{itemize}
 \item Q1: What are the properties of coronal loops when observed with very high spatial resolution from multiple viewing angles?
 \item Q2: What role do small-scale brightenings from the photosphere to the corona play in active region evolution?
 \item Q3: What are the characteristics of coronal rain, and where does it originate on the basis of high-resolution observations from multiple viewing angles?
\end{itemize}
 
We have structured our article as follows: In Sect.~\ref{sect:instruments}, we describe the instruments used from Solar Orbiter and DKIST in these coordinated observations. In Sect.~\ref{sect:obs_strategy}, we present an overview of the observations, from planning to execution. In Sect.~\ref{sect:obs_highlights}, we discuss how the collected observations will help tackle the three open science topics that were outlined in the previous paragraph. Finally, in Sect.~\ref{sect:conclusion}, we provide a summary of these coordinated observations and information on future plans.

\section{Instrumentation}
\label{sect:instruments}

\subsection{Solar Orbiter}

Solar Orbiter carries ten instruments on board, six of which sample remote sensing observations of the solar atmosphere and the inner heliosphere, while the other four collect in situ measurements of the environment local to the spacecraft. Three of the remote sensing instruments provided the data most relevant to the coordinated observations analysed here, namely, 
the Extreme-Ultraviolet Imager \citep[EUI;][]{Rochus2020}, the Spectral Imaging of the Coronal Environment \citep[SPICE;][]{SPICE2020, Fludra2021} instrument, and the Polarimetric and Helioseismic Imager \citep[PHI;][]{Solanki2020}. This section provides a general description of these three instruments. Detailed information about the pointing, field of view (FOV), and plate scale for each instrument during the coordinated observations are presented in Table~\ref{table:solOrb_obs}. The Solar Orbiter data are available from the Solar Orbiter Archive\footnote{ (\url{https://soar.esac.esa.int/soar/}} and from the local instrument archive web pages indicated below.

\subsubsection{Extreme-Ultraviolet Imager}

The EUI instrument observes the solar disc and off disc with three telescopes: the full disc imager (FSI), which observes using two filters, 174 \AA~ (FSI~174) and 304 \AA~ (FSI~304), and two high-resolution imagers (HRIs) that observe the Sun in the Lyman-$\alpha$ passband centred around 1216 \AA~(HRI$_{Ly-\alpha}$) and an extreme ultraviolet (EUV) passband centred around 174 \AA\ (HRI$_{EUV}$), respectively. The HRI$_{Ly-\alpha}$ telescope filter includes chromospheric emission from the Hydrogen Lyman-$\alpha$, whilst the HRI$_{EUV}$ telescope observes the lower coronal emission at a temperature of $\approx$ 1 MK, mainly originating from the \ion{Fe}{IX} and \ion{Fe}{X} lines. In all the figures plotted in this work including EUI data (both FSI and HRI), we use Level-2 data that were provided by the EUI team in the EUI Data Release~6.0 2023-01\footnote{ (\url{https://doi.org/10.24414/z818-4163}}. During the coordinated observations between Solar Orbiter and DKIST, the Sun was observed most of the time (19-23 October 2022) with all EUI telescopes in all filters (FSI~304, FSI~174, HRI$_{EUV}$, HRI$_{Ly-\alpha}$). Exceptions include 18 October 2022, when observations were provided only by FSI~304 and FSI~171, and 24 October 2024, when observations were provided by FSI~304,  FSI~171, and HRI$_{EUV}$. Detailed information about the observations in different wavelength channels are presented in Table~\ref{table:solOrb_obs}.

\subsubsection{Spectral Imaging of the Coronal Environment instrument}\label{sec:spice}

The SPICE instrument is an imaging spectrometer observing the Sun in extreme-ultraviolet and sampling nominally the solar atmosphere from the upper chromosphere to the corona, including flares. SPICE observes the Sun in two wavelength ranges: the short wavelength (SW) range, from 704 to 790 \AA~, and the long wavelength (LW) range, from 973 to 1049 \AA. The temperature range covered by the SPICE instrument spans from around $10^4$ K (\ion{H}{I}) to $2\times10^7$ K (\ion{Fe}{XX}). However, the hot emission in \ion{Fe}{XX} is observed only during the flaring activity. SPICE observes the Sun in one of three modes. These are the traditional sit-and-stare and raster observations as well as the slit mode, where a single wide (30\arcsec) slit is employed. For the sit-and-stare and raster modes, three different slit widths can be selected (2\arcsec, 4\arcsec, 6\arcsec). We used the Level 2 data release v.04\footnote{\url{https://spice.osups.universite-paris-saclay.fr/spice-data/release-4.0/release-notes.html}} for all the SPICE plots in this article. The SPICE data are available directly from the Solar Orbiter Archive. During the Long-Term Active Region SOOP, two types of studies were run: \texttt{SCI\_COMPO-TEST2\_SC\_SL04\_60.0S\_FF} and \texttt{SCI\_DYN-MED\_SC\_SL04\_5.0S\_FF}. The aim of the first study was to observe the spectral lines that can be useful for composition studies \citep[][private communication]{Mzerguat-submitted}. This was done by providing a dense raster using a 4\arcsec-wide slit with an exposure time of 60\,s per scan step and without on-board binning. The composition study was run once every day, for a total of $\approx$3.25h, but always during DKIST observation time. The second study was run for the rest of the day each day, and its aim was to provide spectrograms at a higher cadence (once every $\approx$15') to capture dynamical events happening in the region of interest. It provides a dense raster, also with the 4\arcsec-wide slit, with an exposure time of 5\,s per scan step and without on-board binning.

\subsubsection{Polarimetric and Helioseismic Imager}

The PHI instrument provides maps detailing several physical quantities in the lower solar atmosphere, including the line-of-sight flow velocity and vector magnetic field values (i.e. field strength, inclination, ambiguous azimuth). These components allowed us to obtain the line of sight and transverse components of the photospheric magnetic field. PHI obtains maps of these quantities by measuring the polarisation state of the incoming light with a narrow-band filter at several wavelength positions (including in the nearby continuum) around the photospheric 6173~$\AA$ Fe I line. The PHI instrument comprises two telescopes: the full disk telescope (FDT), which as the name suggests observes the entire solar disc, and the high-resolution telescope (HRT; \citealt{Gandorfer18}), which can be used to observe smaller regions on the solar disc with increased spatial and temporal resolutions. In our plots, we use Level 2 data for the HRT data, which are available in the Solar Orbiter Archive; they are HRT data from the July 2023 data release.\footnote{\url{ https://www.mps.mpg.de/solar-physics/solar-orbiter-phi/data-releases}} Unfortunately, due to an error in the refocusing system during the time interval considered for the Long-Term Active Region SOOP, there are no science-ready FDT data available.

\subsection{The Daniel K. Inouye Solar Telescope}
The telescope DKIST is located at 3084 meters near the summit of Haleakal\={a} on the island of Maui, which is 10 hours behind Universal time (UT). During Cycle 1 of the OCP, the three first-generation instruments available were as follows: the Cryogenic Near-Infrared Spectro-Polarimeter \citep[CryoNIRSP:][]{Fehlmann2023}; the Visible Broadband Imager \citep[VBI;][]{Woeger2021}; and the Visible Spectro-Polarimeter \citep[ViSP;][]{DeWijn2022}. In the following section, we give a brief summary of the three instruments. More detailed information about these experiments is included in Table~\ref{table:dkist_obs}. For coordinating with Solar Orbiter, two separate experiments were defined for DKIST. The first made use of Cryo-NIRSP for off-limb observations, experiment ID (EID) Cycle 1 Proposal \#122, or EID\_1\_122, whereas the second experiment combined VBI and ViSP for on-disc observations (EID\_1\_123). The data returned by VBI and ViSP (EID\_1\_123) are publicly available at the DKIST Data Centre Archive.\footnote{\url{https://dkist.data.nso.edu/}} The CryoNIRSP data (EID\_1\_122) is not yet available on the archive due to challenges in calibrating their early Cycle 1 data.

\subsubsection{Cryogenic Near-Infrared Spectro-Polarimeter}

The Cryo-NIRSP instrument consists of a slit-grating-based spectrograph alongside a narrow-band context imager. The spectrograph can sequentially observe spectral lines between 1 and 4 microns. In particular, it can target forbidden emission lines that are magnetically sensitive in the solar corona. These lines are sensitive to plasma at coronal temperatures and allow us to study the dynamics and evolution of the magnetic field, plasma flow, and density in this region of the solar atmosphere. Importantly, Cryo-NIRSP cannot currently operate in coordination with the other instruments at DKIST and must therefore be used alone. We note that Cryo-NIRSP was used exclusively to collect observations from above the solar limb. The first results obtained using the CryoNIRSP instrument are described in \citet{Schad2023, Schad2024a, Schad2024b}. The plots of Cryo-NIRSP data are provided by the Cryo-NIRSP instrument team prior to data publication on the DKIST data archive.

\subsubsection{Visible Broadband Imager}
The VBI instrument \citep{Woeger2021} provides high-cadence imaging of the lower solar atmosphere at the highest spatial resolution achievable by DKIST. Observations with the VBI can be collected either to optimise spatial coverage (e.g. by mosaicaing and sampling the entire $2$'$\times2$' optical FOV of the VBI) or temporal cadence (e.g. by collecting observations at only a subfield of the VBI). The VBI operates using two distinct channels: the VBI-Blue and the VBI-Red, which can be used in isolation or in combination. Both of these channels make multiple spectral pre-filters available that can be used to observe specific wavelength regions and probe different layers of the solar atmosphere ranging from the photosphere to the chromosphere. Two large-format detectors ($4k\times 4k$, mounted on a two-axis linear stage) allow for critical sampling of the diffraction limit at the shortest wavelength of the respective channel and provide a central FOV of $45\arcsec\times 45\arcsec$ and $69\arcsec\times 69\arcsec$ for the blue and red channels, respectively. Essentially, VBI allows for detailed analysis of the dynamics and evolution of small-scale features in the solar atmosphere. Furthermore, these images provide context for other instruments. In combination, the VBI-Blue and VBI-Red channels allow spectral coverage as well as cadence to be balanced independently from each other (e.g. by operating with several pre-filters in one channel and only a single pre-filter in the other).

\subsubsection{Visible SpectroPolarimeter}

The ViSP instrument \citep{DeWijn2022} is a traditional slit-scanning grating-based spectropolarimeter that is able to observe up to three spectral channels within the spectral range of 3800 to 9000 \AA~using three distinct `arms' simultaneously.
Each arm consists of imaging optics optically downstream from a common diffraction grating, the combination of which can be angularly configured to sample different wavelengths and diffraction orders.
Using a ten-state modulation sequence, ViSP can determine all four polarisation states of the incoming light (Stokes I, Q, U, and V), and, as such, it can be used to infer properties of the plasma and magnetic field in the lower solar atmosphere. ViSP is able to perform raster or sit-and-stare observations depending on the scientific need, with a maximum FOV of around $120''\times75''$.

\section{Observing strategy}\label{sect:obs_strategy}

\begin{figure*}[!htb]
\center{\includegraphics[width=\textwidth]
{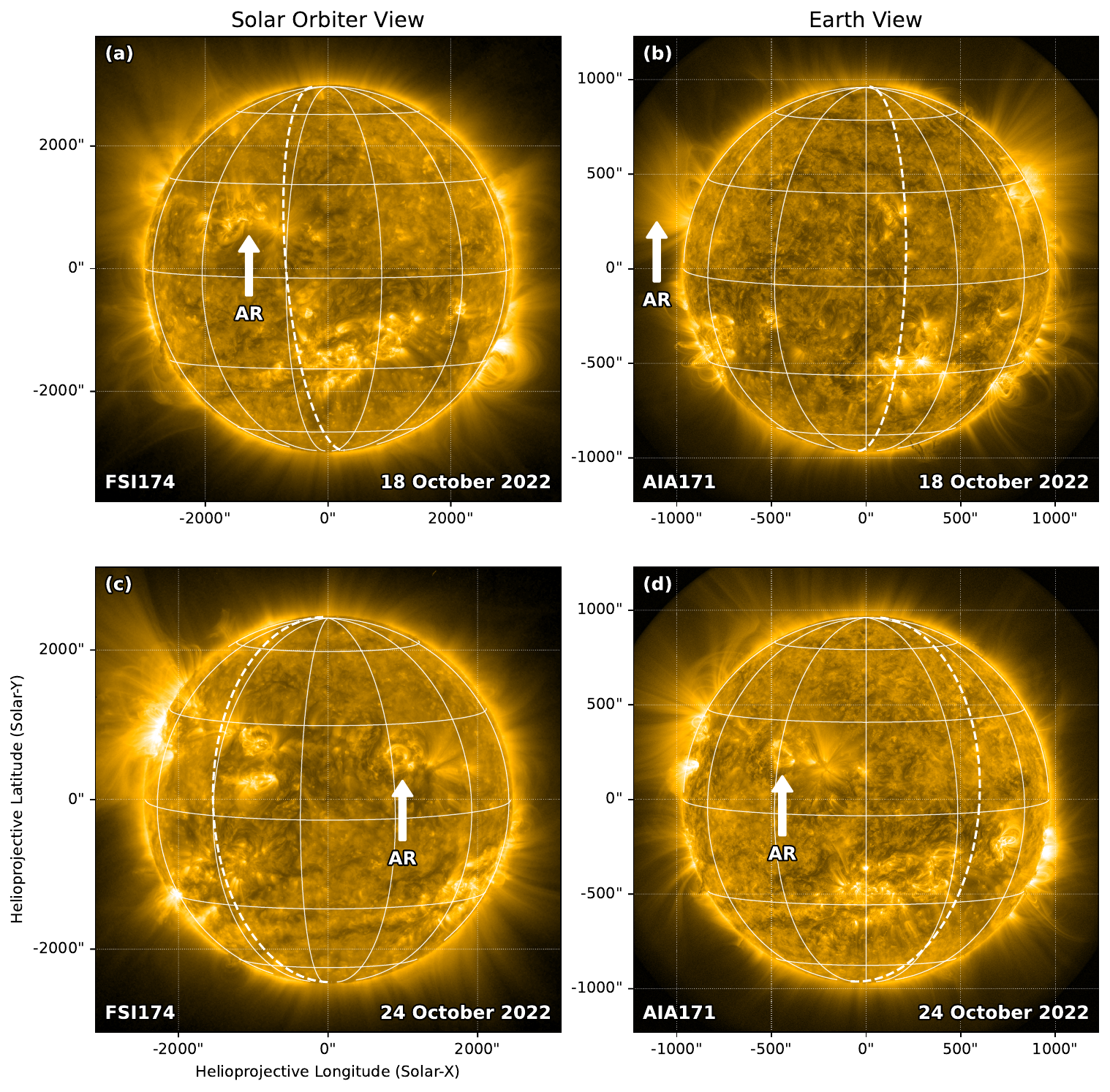}
\caption{Active region (highlighted by the white arrow) selected by the coordinated observations as visible from the Solar Orbiter perspective, as collected by the EUI/FSI $174$\AA~filter (left column), and the Earth perspective, as collected by the SDO/AIA $174$\AA~filter (right column). The top row shows a plot of the Sun on the first day (18 October) of coordination, whilst the bottom row shows a plot of the Sun on the last day (24 October) of coordinated observations. The dashed line highlights the position of the limb as seen from a perspective of the coordinated instrument. The overplotted latitude and longitude grids are in heliographic Stonyhurst coordinates.}
\label{fig:ar_overview_position}}
\end{figure*}

\begin{figure}[!htb]
\center{\includegraphics[width=0.48\textwidth]
{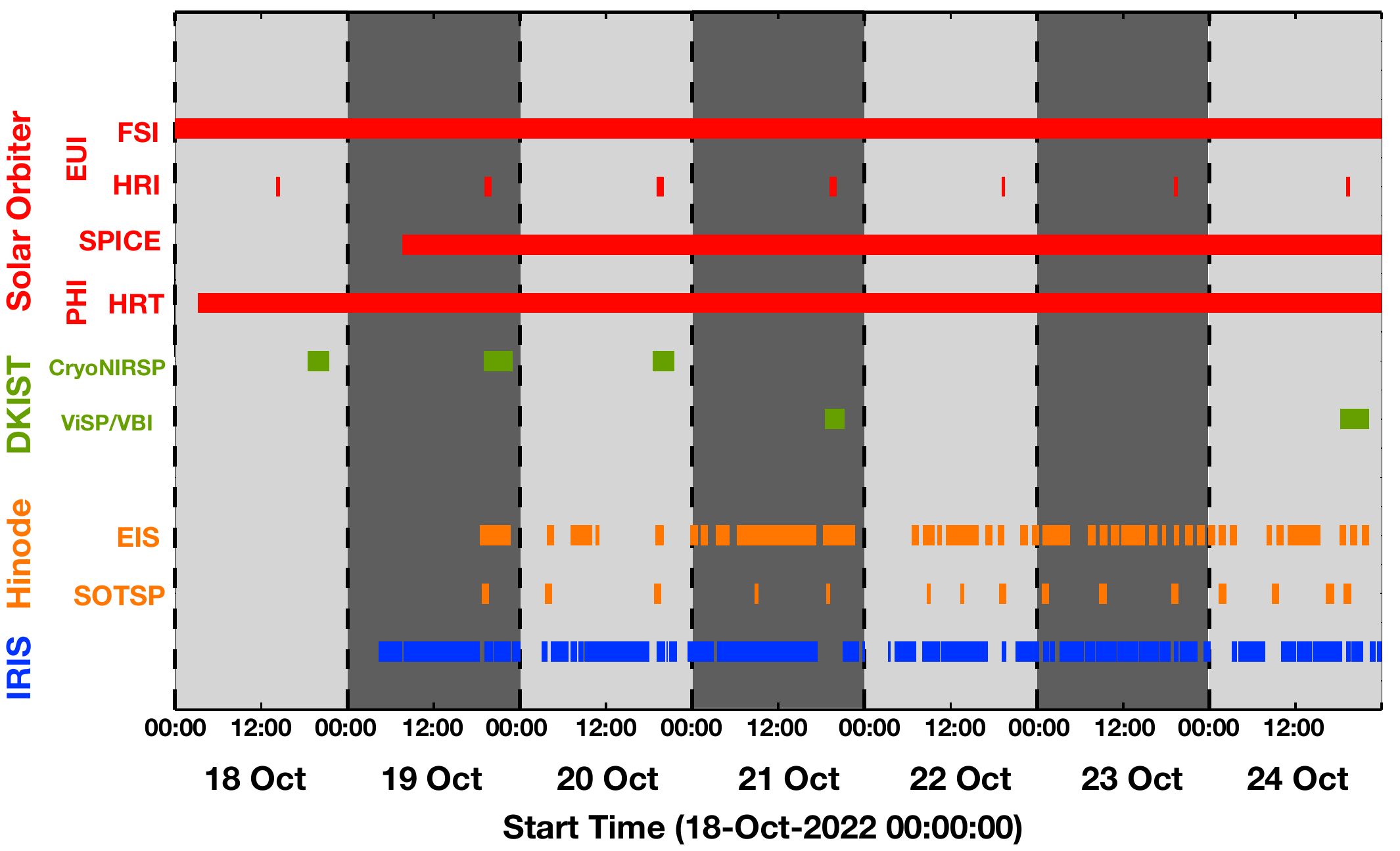}
\caption{Periods of active region observations for selected instruments on board Solar Orbiter and DKIST as well as IRIS and Hinode. The Solar Orbiter PHI instrument provided one image per hour during the entire period.}
\label{fig:overview_observation_time}} 
\end{figure}

\begin{table}

\caption{Position of Solar Orbiter with respect to the Earth and Sun.} 

\label{table:obs_setup} 

\small 
\begin{tabular}{ll c c p{8cm}} 

\hline\hline 
{Time} &{Ang. sep. [deg]} & {Distance [au]} \\
\hline\hline 
18 October 2022 & 76.66 & 0.32\\
\hline 
19 October 2022 & 71.52 & 0.33\\
\hline
20 October 2022 & 66.74 & 0.34\\
\hline 
21 October 2022 & 62.33 & 0.36\\
\hline 
24 October 2022 & 51.13 & 0.39\\
\hline 

\end{tabular}
\tablefoot{This table contains the Solar Orbiter-Sun-Earth angle (centre column) as well as the distance between Solar Orbiter and the Sun (right column). The values are presented at 19:05 UT for the days listed in the table.}
\end{table}

\subsection{Target selection}
\label{obs_target}

Solar Orbiter reached its perihelion during its 6-month orbit on 12 October 2022 at 19:12~UT, reaching a distance of only 0.293 au from the Sun. On 16 October 2022, Solar Orbiter, the Sun, and Earth were in quadrature. After that, the angular separation between the Solar Orbiter and Earth with respect to the Sun began to decrease, reaching 76.66$^\circ$ on the first day of coordinated observations with DKIST (18 October) before decreasing to 51.13$^\circ$ on the last day of coordinated observations with DKIST (24 October). In Fig.~\ref{fig:ar_overview_position}, we plot the full solar disc as observed from the perspective of the Solar Orbiter spacecraft (left column) and the Solar Dynamics Observatory's Atmospheric Imaging Assembly's (SDO/AIA; \citealt{Lemen2012}) $171$~\AA\ filter (right column) on 18 (top row) and 24 (bottom row) October. Detailed information on the angular separation and distance between Solar Orbiter and the Sun can be found in Table~\ref{table:obs_setup}. 

Target selection for Solar Orbiter and co-observing with other telescopes is  primarily at the discretion of the SOOP coordinators (science leaders), a group of scientists who ensure that the observational routines being run by the on-board instruments are optimised in order to answer the scientific goals of their particular SOOP. Between 18 and 24 October 2022, Solar Orbiter was conducting an instance of the Long-Term Active Region SOOP with a heavy focus on the decay phase of active region lifetime. The decay phase is a phase of an active region lifetime that is characterised by the gradual weakening and finally disappearance of magnetic fields. A target appearing in the Solar Orbiter FOV a few days prior to the start of the SOOP was identified as AR13110, a relatively dynamic active region observed 15 days prior by Earth assets. As no synoptic magnetograms were available during the pointing decisions, the views from Earth, complemented by the early views of Solar Orbiter instruments, gave some insight into the evolution of the active region. Due to its location on the eastern limb, as seen from Solar Orbiter, this active region was an ideal target for long-term tracking by Solar Orbiter, and this was followed a few days later by Earth assets. 
On the first day (18 October 2022) of the coordinated observation campaign, the active region was therefore observed in the northeast part of the solar disc with Solar Orbiter (Fig.~\ref{fig:ar_overview_position}a). From the Solar Orbiter perspective, the coordinates on the active region centre are [-1400\arcsec, 800\arcsec] in the helioprojective coordinate system, 18 October 2022 at 19:00 UT. At this time, ground-based and Earth-orbiting instruments showed that only the upper part of the coronal loops extended significantly in the east direction from the limb of an active region (Fig.~\ref{fig:ar_overview_position}b). During the seven days of coordinated observations, the observed active region evolved and presented large-scale coronal loop evolution, small-scale structure evolution, and coronal rain. However, no flare stronger than the B-class was observed during that time. By the time the active region rotated fully into Earth's view, magnetograms provided by Earth's assets showed a highly decayed region, with no provided NOAA number.
On the last day of coordinated observations (24 October 2022), Solar Orbiter observed the same active region at the northwestern part of the solar disc (Fig.~\ref{fig:ar_overview_position}c). From Earth-orbiting satellites and ground-based observatories, the active region was observed in the north-eastern quadrant of the solar disc (Fig.~\ref{fig:ar_overview_position}d). 

In this article, we split the coordinated observations into two distinct parts due to the different positions of the active region on the solar disc as observed from Earth. The reason for this is that the active region position implies the use of different DKIST instruments. During the first part of the coordinated observations, we aimed to observe the active region close to the limb from Earth's perspective (18-20 October 2022), and thus, CryoNIRSP observations were collected. The second part focused on on-disc observations (21-24 October 2022), and therefore, VBI and ViSP observations were collected. We discuss the two parts of the observations in Section~\ref{sec:off_limb} and Section~\ref{sec:on_disk}, respectively.

In addition to Solar Orbiter and DKIST, these observations were also coordinated with the Earth orbiting satellites Hinode \citep{Kosugi2007} and the Interface Region Imaging Spectrograph \citep[IRIS;][]{DePontieu2014}, which obtained spectroscopic observations of the chromosphere, transition region, and the solar corona. A detailed investigation of the plasma composition evolution during the observations with Solar Orbiter/SPICE and Hinode/EIS can be found in \citet[][private communication]{Mzerguat-submitted}, and \citet{Varesano2024, Varesano2025} Here, we provide information on the timings of these data for potential users. The observation periods of the remote-sensing instruments on board Solar Orbiter, DKIST, Hinode, and IRIS are shown in Fig.~\ref{fig:overview_observation_time}.

\subsection{Active region off-limb observation from Earth view }
\label{sec:off_limb}

\begin{figure*}[!htb]
\center{\includegraphics[width=\textwidth]
{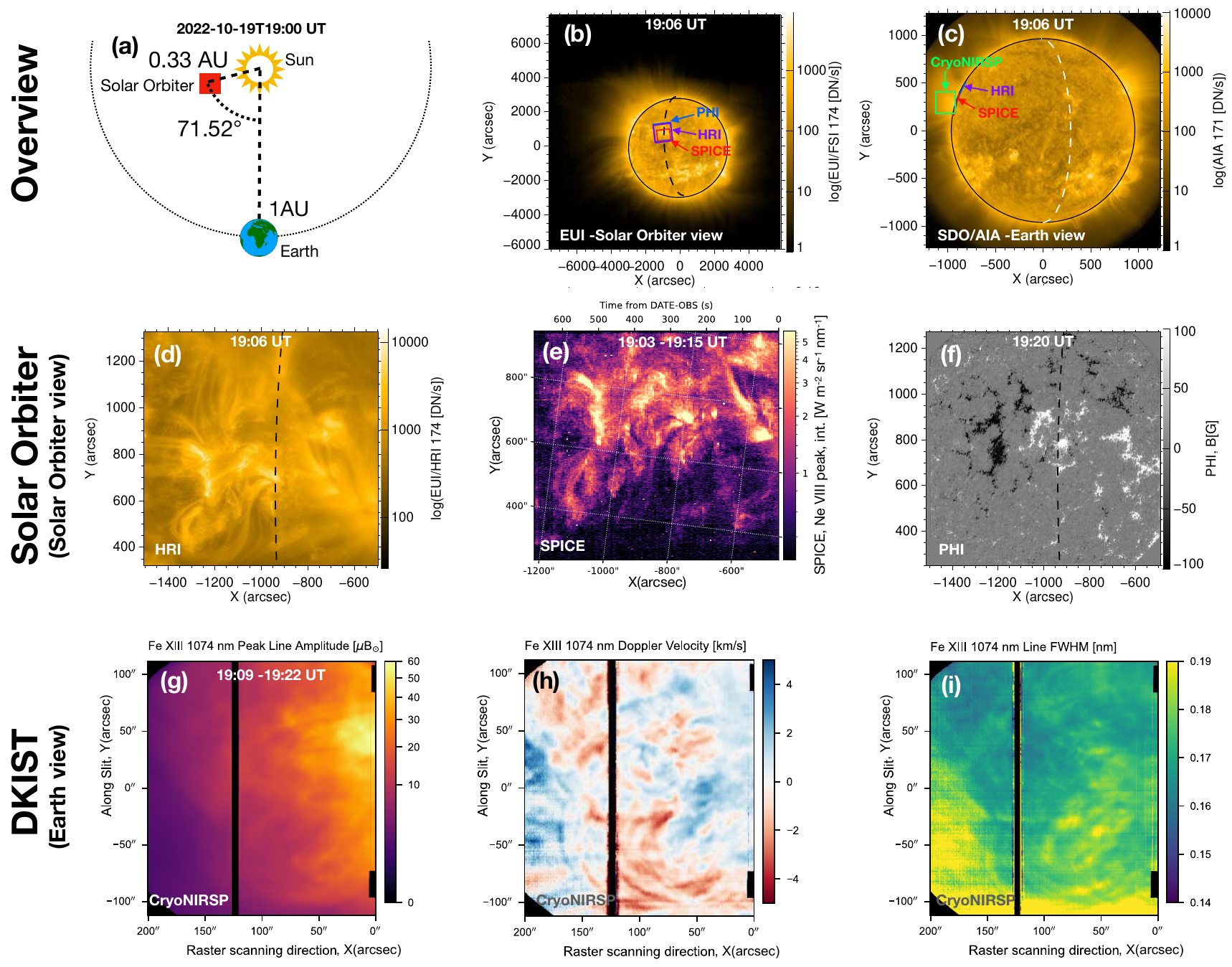}
\caption{Overview of coordinated observations provided by Solar Orbiter and DKIST on 19 October 2022. (a) Relative positions of the Solar Orbiter spacecraft and the Earth. (b) Sun observed with Solar Orbiter. The coloured boxes show the fields-of-view of PHI/HRT (blue), SPICE (red), and the EUI/HRI (violet). (c) Sun observed with SDO/AIA with the same view as from DKIST with the CryoNIRSP FOV overlaid (green). (d) EUI/HRI image shows the coronal emission with high resolution.(e) SPICE raster map showing the transition region emission intensity peak of the \ion{Ne}{viii} spectral line obtained during dynamics observations \texttt{SCI\_DYN-MED\_SC\_SL04\_5.0S\_FF}. (f) PHI/HRT magnetograms showing the line-of-sight magnetic field in a range between -100 and 100 G. (g) DKIST CryoNIRSP observation showing the coronal loops as observed using the \ion{Fe}{xiii} line. (h) Corresponding Doppler velocity maps showing a speed of around $\pm$4 km/s. (i) Corresponding FWHM map. The dark vertical band in panels (g), (h), and (i) represents a region where clouds impacted the observations.
}
\label{fig:overview_fov_19oct}} 
\end{figure*}

The first phase of the coordinated observations took place from 18 October to 20 October 2022. During this time, the EUI/FSI chromospheric (FSI 304\AA) and coronal (FSI 171\AA) filters were used to obtain full disc maps at 10-30 min intervals, PHI/HRT provided high-resolution magnetograms at a cadence of 1 hour, and SPICE provided spectroscopic raster scans over the active region continuously (see Section~\ref{sec:spice}). High-resolution imaging data collected by EUI/HRI were obtained in distinct 30-60 minute windows using both the Ly-alpha (1216\AA) and EUV (171\AA) telescopes. These high-resolution EUI/HRI observations were first obtained on 19 October 2022. 

The telescope DKIST provided off-limb observations (Experiment EID\_1\_122) with CryoNIRSP only. The CryoNIRSP raster observed the active region and sampled the solar corona from just above the limb to 1.24 solar radii with fields of view of 167$\times$145~Mm (230$\times$200\arcsec) for the spectrograph and 72.5$\times$72.5~Mm,(100$\times$100\arcsec) for the context imager. The CryoNIRSP slit was located parallel to the solar limb. The spectral data for two \ion{Fe}{xiii} lines was sampled in the spectroscopic-only mode (in order to provide information about the coronal intensity and Doppler velocity, which provide information about coronal plasma topology and flow, respectively) and the full-Stokes polarimetric mode (in order to infer information about the coronal magnetic field structuring). The intensity ratio of this \ion{Fe}{xiii} line pair (10746 and 10798~\AA) provides a diagnostic of the electron density. The slit width of the CryoNIRSP instrument is $0.5\arcsec$ wide, whilst the slit length is ${\sim}200\arcsec$, with $0.12\arcsec$ per pixel sampling. The spectral resolution, $R$, is ${\sim}40000$ in this mode. We used a combination of spectroscopic-only and fully polarimetric observing programs (OPs), which included both the spectrograph and context imager components of CryoNIRSP, namely,
\begin{itemize}
 \item OP1: Telescope centre pointing at 1.08 solar radii with the slit parallel to the limb, spectroscopic-only 400-step rasters in \ion{Fe}{xiii} 10740 and \ion{Fe}{xiii} 10790~\AA, and context imager data at He I 10830~\AA. 
 \item OP2: Telescope centre pointing at 1.08 solar radii with slit parallel to limb, full-Stokes polarimetric 100-step rasters in \ion{Fe}{xiii} 10740~\AA, and context imager He I 10830~\AA.
 \item OP3: Telescope centre pointing at 1.08 solar radii with slit parallel to limb, deep full-Stokes polarimetric coarse ten-step rasters for magnetic field measurements off-limb, spectrograph \ion{Fe}{xiii} 10740~\AA, and context imager He I 10830~\AA.
\end{itemize}

Figure~\ref{fig:overview_fov_19oct}a shows the configuration of Solar Orbiter, the Sun, and the Earth on the first (19 October 2022) day of high-resolution coordinated observations (i.e. when the EUI/HRI instrument coordinated with DKIST). Clearly, the selected targeted was off-limb from the viewpoint of  DKIST, whilst it was close to disc centre from the Solar Orbiter viewpoint. The observations obtained on 18 October 2022 (without EUI/HRI support) are promising for future studies of the coronal loop properties (Q1) and characteristics of coronal rain and its origin (Q3). The data from 19 and 20 October 2022 (with EUI/HRI support) are promising for the analysis of coronal loop systems in active regions (Q1), the role of small-scale structures in active regions (Q2), and coronal rain (Q3). A detailed description of the first part of the coordinated observations is given below. We discuss each day separately.

\begin{figure*}[!htb]
\center{
\includegraphics[width=0.95\textwidth]{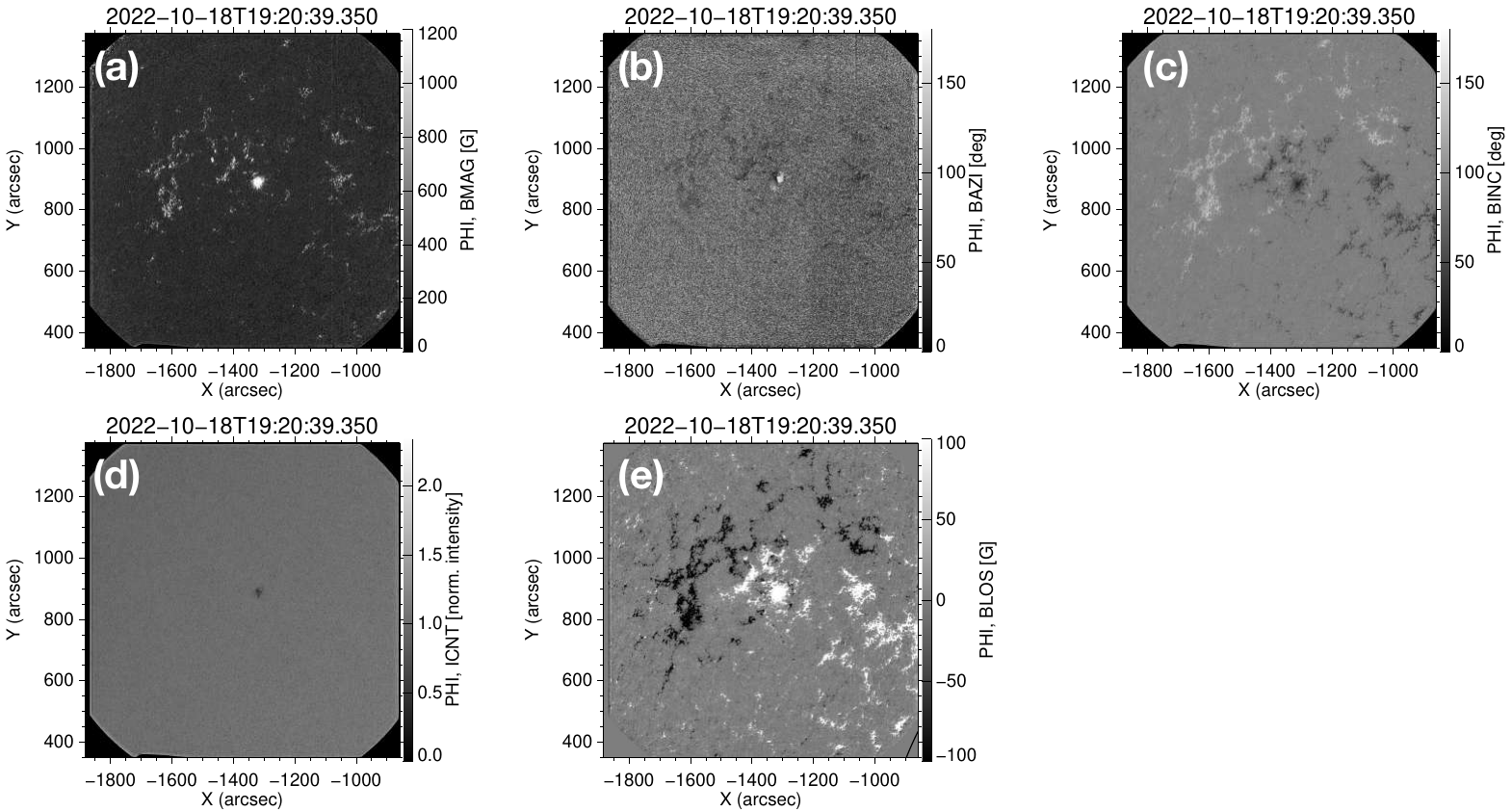} 
\caption{Magnetic field observations captured by SO/PHI on 18 October 2022 at 19:20 UT. The following parameters of the magnetic field vector are illustrated in the accompanying figure: (a) magnetic field strength, (b) ambiguous magnetic field azimuth, and (c) magnetic field inclination. There are also maps of (d) continuum intensity and (e) the line of sight of the magnetic field. In all panels, the roundish border at the corners is the field stop of HRT.}\label{fig:fig_mag}}
\end{figure*}

\subsubsection{18 October 2022}

On 18 October, the Solar Orbiter target was slightly behind the solar limb as observed from Earth. On this day, SPICE, PHI/HRT, and EUI/FSI obtained images of the active region as it evolved through time. The EUI/FSI observations clearly show an active region with numerous coronal loops, whilst time series of the PHI/HRT line-of-sight magnetograms show continuously evolving small-scale magnetic structures surrounding the large-scale magnetic patches (Fig.~\ref{fig:fig_mag}, see Movie-PHI.mp4). Although the target active region was behind the limb as observed from Earth, the coronal loops evident in EUI/FSI were visible off the solar limb. As such, DKIST conducted tests of the CryoNIRSP observations on that target. These tests were successful and allowed for a full set of science observations to be acquired. The OPs of Experiment EID\_1\_122 were executed multiple times (in the order: OP1, OP2, OP1, OP1, OP3, OP1) between 18:30 UT and 21:22 UT. Combined, the EUI/FSI and CryoNIRSP observations show slowly evolving coronal loops in the active region (Q1), which can be studied in detail using the stereoscopic measurements obtained through this coordination. During the coordinated observation, the spatial area occupied by the active region is almost the same.

\subsubsection{19 October 2022}
On 19 October 2022, the angular separation between the Earth and Solar Orbiter was 71.52$^\circ$ with respect to the Sun, and Solar Orbiter was located at 0.33 au from the Sun (Fig.~\ref{fig:overview_fov_19oct}a). On this date, the active region that was tracked by Solar Orbiter was close to the disc centre from its viewpoint (Fig.~\ref{fig:overview_fov_19oct} b) and was very close to the solar limb as observed from Earth (Fig.~\ref{fig:overview_fov_19oct}c). From Solar Orbiter, SPICE, PHI/HRT, EUI/FSI, and EUI/HRI observations were collected. Specifically, high-cadence and high-resolution EUI/HRI observations were collected between 19:00 UT and 20:00 UT. The active region shows extended coronal loops in the EUI/HRI observations
(Fig.~\ref{fig:overview_fov_19oct}d), brightenings in the transition region spectroscopic SPICE raster map presented peak intensity in \ion{Ne}{viii} (Fig.~\ref{fig:overview_fov_19oct}e), and bipolar plage topology in the underlying photospheric magnetic field (Fig.~\ref{fig:overview_fov_19oct}f). Between 19:09 UT and 23:29 UT, DKIST CryoNIRSP successfully obtained numerous complete observations using the pre-defined OPs (in the order: OP1, OP1, OP2, OP3, OP1, OP1, OP2, OP1, OP1). The DKIST CryoNIRSP intensity map of the \ion{Fe}{xiii} line (Fig.~\ref{fig:overview_fov_19oct}g) shows an evolving loop system above the solar disc. Fitting Gaussian functions to the spectral profile of the \ion{Fe}{xiii} line allowed us to derive a Doppler velocity map (Fig.~\ref{fig:overview_fov_19oct}h) across the coronal loops. The coronal plasma observed in the loop system moves with the small velocity of $\pm$ 4~km/s. These stereoscopic observations revealed both coronal loop evolution (Q1) and coronal rain (Q3). Detailed descriptions of these are presented in Section~\ref{sect:obs_loop} and Section~\ref{sect:coronal_rain}, respectively.

\subsubsection{20 October 2022}

On 20 October 2022, the observation setup and the contributing instruments were the same as on 19 October 2022. The EUI/HRI observations were obtained between 19:00~UT and 20:00~UT, whilst the CryoNIRSP off-limb observations were collected in the time frame between 18:41 UT and 21:36 UT. On this date, fewer OPs were run, but several successful experiments were still conducted (in the order: OP1, OP1, OP2, OP3, OP1). These observations show a reconfiguration of large-scale coronal loop substructures (Q1), the evolution of numerous small-scale structures (Q2), and coronal rain (Q3).

\subsection{Active region on disc observations as seen from Earth}
\label{sec:on_disk}

\begin{figure*}[!htb]
\center{\includegraphics[width=\textwidth]
{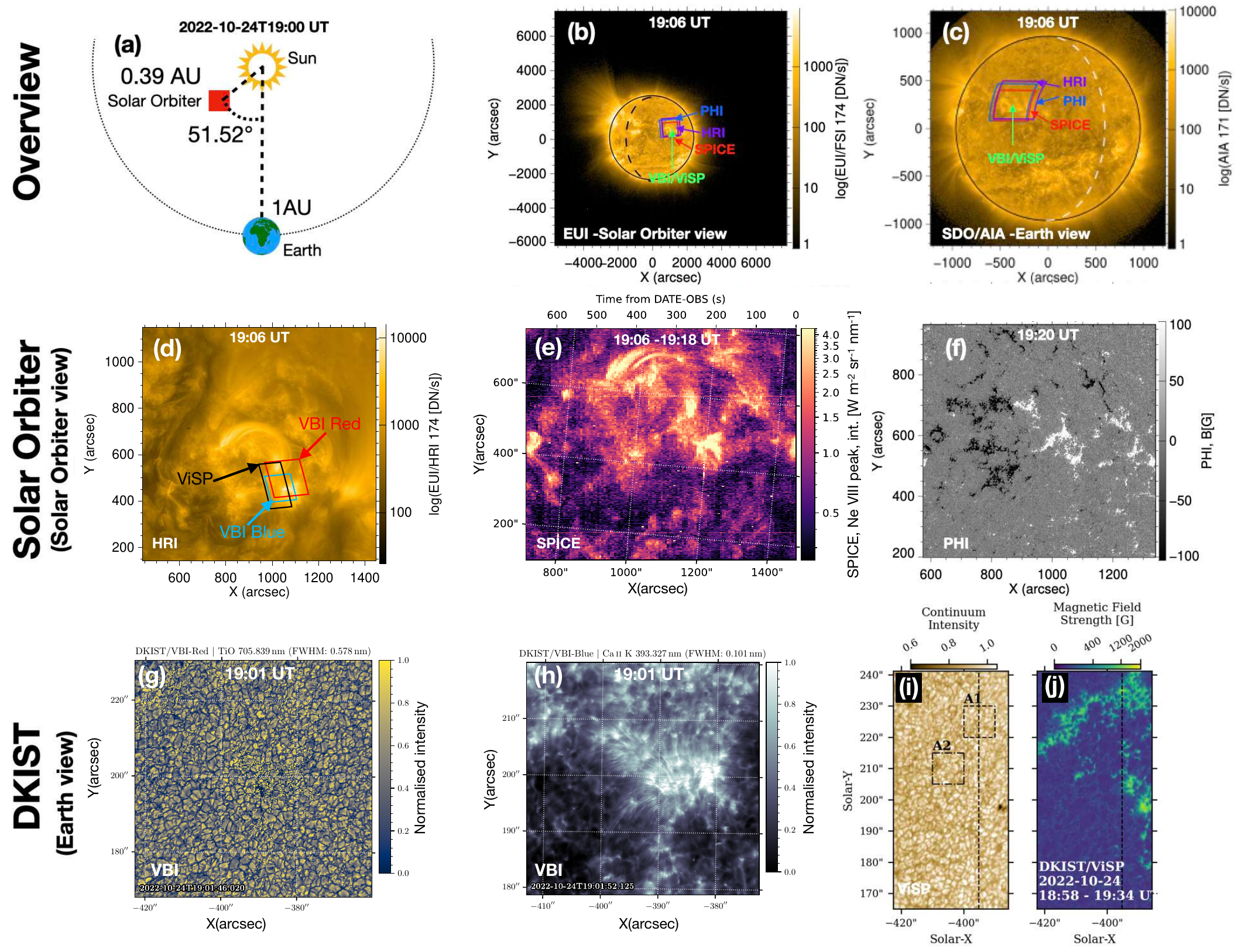}
\caption{Overview of the coordinated observations provided by Solar Orbiter and DKIST on 24 October 2022. (a) Position of the Solar Orbiter spacecraft with respect to the Earth. (b) Sun as observed from the vantage point of Solar Orbiter. The coloured boxes again outline the fields of view of the coordinating instruments. (c) Sun observed with SDO/AIA, the same view as from DKIST, with the fields of view of the Solar Orbiter instruments overlaid. (d) HRI image showing the coronal emission with a high-resolution. The overlaid boxes outline the fields of view of the DKIST instruments: Red is for the VBI-Red FOV, blue is for the VBI-Blue FOV, and black is for the ViSP raster FOV. (e) SPICE raster map showing the transition region emission peak intensity of the \ion{Ne}{viii} spectral line. (f) PHI magnetograms showing the magnetic field in range between -100 and 100 G. (g) DKIST VBI-Red band image showing the photospheric granulation in the TiO spectral line. (h) DKIST VBI-Blue band image showing the chromosphere in CaII K spectral line. (i) DKIST ViSP data showing the photospheric granulation observed in continuum. (j) Corresponding absolute magnetic field strength.}
\label{fig:overview_fov_24oct}} 
\end{figure*}

The second phase of these coordinated observations took place on 21 and 24 October 2022. On these dates, the active region that was being tracked was now on disc as seen for both Solar Orbiter and Earth assets, being located at a mid-distance between the disc centre and the limb in the east solar hemisphere from the viewpoint of Solar Orbiter (Fig.~\ref{fig:ar_overview_position}c) and in the west solar hemisphere from the viewpoint of Earth (i.e. DKIST; Fig.~\ref{fig:ar_overview_position}d). The observations obtained by Solar Orbiter were similar to those collected on the previous days of coordinated observations with one slight difference: the PHI/HRT observations from 24 October were cropped to 1536$\times$1536 px on board. 

The main difference between the off-limb and on-disc (as observed from Earth) phases of the coordinated observations was the optical setup of DKIST. For the on-disc observations, DKIST collected imaging and spectro-polarimetric observations with the VBI and ViSP instruments, respectively. During this time, the DKIST Facility Instrument Distribution Optics was set to Configuration 2b (as defined in the Cycle 1 Proposal Call\footnote{\url{https://nso.edu/dkist/proposal-call-1/}}), and as such, we attempted to optimise our OPs accordingly. For VBI, this consisted of combining mosaics and sit-and-stare time series using a variety of photospheric and chromospheric pre-filters. For ViSP, we conducted rasters across the active region sampling photospheric lines. Below, we present the detailed setup of the ViSP and VBI instruments during these experiments. The experiment identifier for these observations (which can be used to search for them in DKIST data centre) is {EID\_1\_123}.\footnote{\url{https://dkist.data.nso.edu/?proposalId=pid_1_123}}

During these ViSP observations, only Arm 1 was enabled. This arm observed a spectral range of 12.5 \AA\ centred at 6301.75~\AA with spectral order nine. This spectral region includes both of the Zeeman sensitive Fe I doublet at 6301.5 ($g_{eff}=1.67$) and 6302.5~\AA\ ($g_{eff}=2.5$). These data allow the user to infer information about the photospheric plasma and magnetic field. The selected entrance slit was 0.0536\arcsec\ wide, with this slit then being rastered across the solar surface using 0.0536\arcsec steps. The FOV along the slit for Arm 1 is 76\arcsec, with a spatial sampling of 0.029\arcsec per pixel. An exposure time of 4 ms in 10 modulation states was repeated 12 times for a total of 480 ms of integration per slit position. Raster maps of up to 1000 slit positions were acquired within a single OP, with numerous repeats being acquired as described below. The runtime of the requested OP is defined by the map time of the ViSP, which was dependent on the number of raster positions. We confirmed a spectral resolution of $R\sim250000$ by analysing the width of telluric absorption lines in the same bandpass.

\begin{figure*}[!htb]
\includegraphics[width=0.95\textwidth]{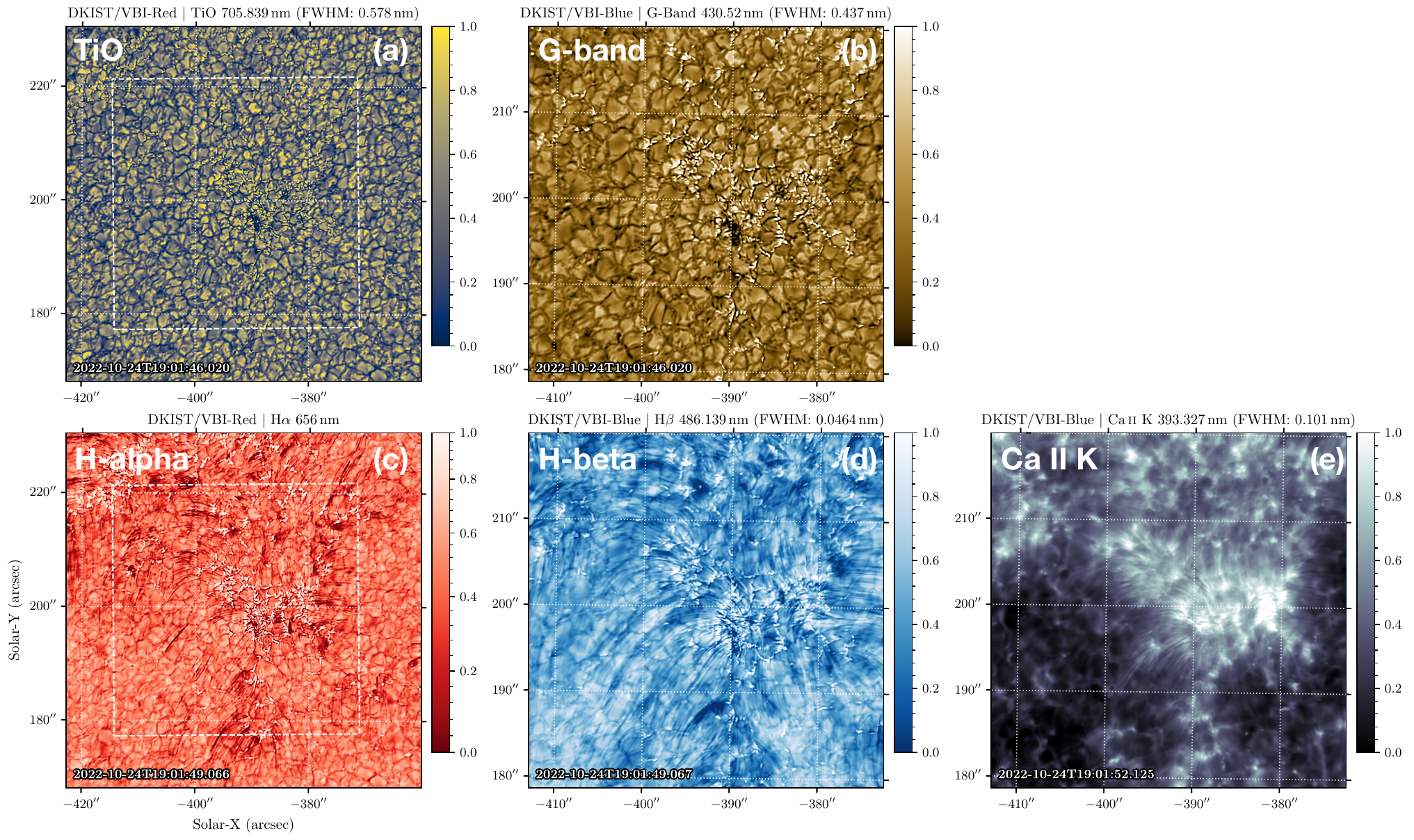} 
\caption{Overview of the VBI observations (both blue and red) obtained by DKIST on 24 October 2022. The top row shows a plot of the solar atmosphere observed using the photospheric TiO (a; VBI-red) and G-band (b; VBI-blue) filters. The bottom row shows a plot of the solar atmosphere observed using the chromospheric H-alpha (c; VBI-red), H-beta (d; VBI-blue), and Ca II k (e; VBI-blue) filters.  The H-alpha filter was found to be off-band for that period, and this is why the chromospheric features are more visible in H-beta than in H-alpha. The white dash line presented in the VBI-Red images (a,c) highlights the VBI-Blue.}
\label{fig:dkist_vbi}
\end{figure*}

In Fig.~\ref{fig:overview_fov_24oct}, we plot a snapshot of the coordinated observations for reference. Fig.~\ref{fig:overview_fov_24oct}a shows the configuration of the Solar Orbiter spacecraft, the Sun, and the Earth on the final day (24 October 2022) of coordinated observations. At this time, the angular separation between Solar Orbiter and Earth with respect to the Sun had decreased to around 51$^\circ$, whilst the distance between Solar Orbiter and the Sun had increased to 0.39\,au (Fig.~\ref{fig:overview_fov_24oct}a); the changes are with respect to the first date of coordinated observations. The active region at this time was located close to the solar disc centre from both Solar Orbiter and Earth (Fig.~\ref{fig:overview_fov_24oct}b,c), with a clear loop system detectable in the EUI/HRI (Fig.~\ref{fig:overview_fov_24oct}d) and SPICE (Fig.~\ref{fig:overview_fov_24oct}e) observations. The fields of view of the DKIST instruments have been overlaid on the EUI/HRI image for reference. A clear bipolar structure is evident within the PHI/HRT date (Fig.~\ref{fig:overview_fov_24oct}f); however, no large-scale structuring (e.g. a sunspot or pore) is apparent in the photospheric or chromospheric imaging collected by the VBI instrument (Figs.~\ref{fig:overview_fov_24oct}g,h, respectively). The ViSP maps (Figs.~\ref{fig:overview_fov_24oct}i,j) detail the presence of strong magnetic fields (up to 2 kG) within this bipolar region. A zoom in on these observations is shown in Fig~\ref{fig:dkist_visp} of \autoref{sec:appendix_visp}. 

During EID\_1\_123, we used both the VBI-Blue and VBI-Red channels to sample images of the solar atmosphere (Fig.~\ref{fig:dkist_vbi}). With VBI-Red, we observed using the TiO (Fig.~\ref{fig:dkist_vbi}a)  and H-$\alpha$ (Fig.~\ref{fig:dkist_vbi}c) filters, whilst with VBI-Blue, we observed using the G-band (Fig.~\ref{fig:dkist_vbi}b), H-$\beta$ (Fig.~\ref{fig:dkist_vbi}d) and Ca II K (Fig.~\ref{fig:dkist_vbi}e) filters.  For both VBI-Blue and VBI-Red, we used a three-step OP consisting of (1) a single mosaic over the entire optical FOV (which is comparable in size to the EUI FOV) followed by (2) time series speckle imaging at the centre field with default settings before (3) a second single mosaic covering the entire optical FOV. For VBI-Blue, the time series collected during step two had a cadence of around $9$ s. For VBI-Red, the time series had a cadence of around $6$ s. Below, we provide a short summary of the observations on both the 21 and 24 October 2022 for completeness.

\subsubsection{21 October 2022}

On 21 October, sustained observations of the active region belonging to the Long-Term Active Region SOOP continued with Solar Orbiter instruments. In order to facilitate these coordinations, EUI/HRI observations were scheduled for a time when DKIST could also perform an observation. These occurred between 19:00 UT and 20:00 UT. DKIST observed using the ViSP and VBI instruments between 18:44 UT and 21:17 UT. This allowed the ViSP instrument to obtain two complete rasters using full-Stokes polarimetry covering an FOV of $39\times54$ Mm$^2$.

\subsubsection{24 October 2022}

On 24 October, the final day of the coordinated observations between Solar Orbiter and DKIST took place. Both facilities conducted observations using the same setup as on 21 October 2022. On this day, though, the EUI/HRI high-resolution and high-cadence observations were obtained for only half an hour, between 19:00~UT and 19:30~UT. DKIST observed between 18:15~UT and 22:15~UT, during which time four ViSP full-Stokes polarimetric maps were obtained. The coverage of each raster in the $x$-direction was $19$ Mm, $28$ Mm, $27$ Mm, and $39$ Mm, respectively, whilst the coverage in the $y$-direction was $54$ Mm in each case.
 
\begin{figure*}[!htb]
\includegraphics[width=0.95\textwidth]{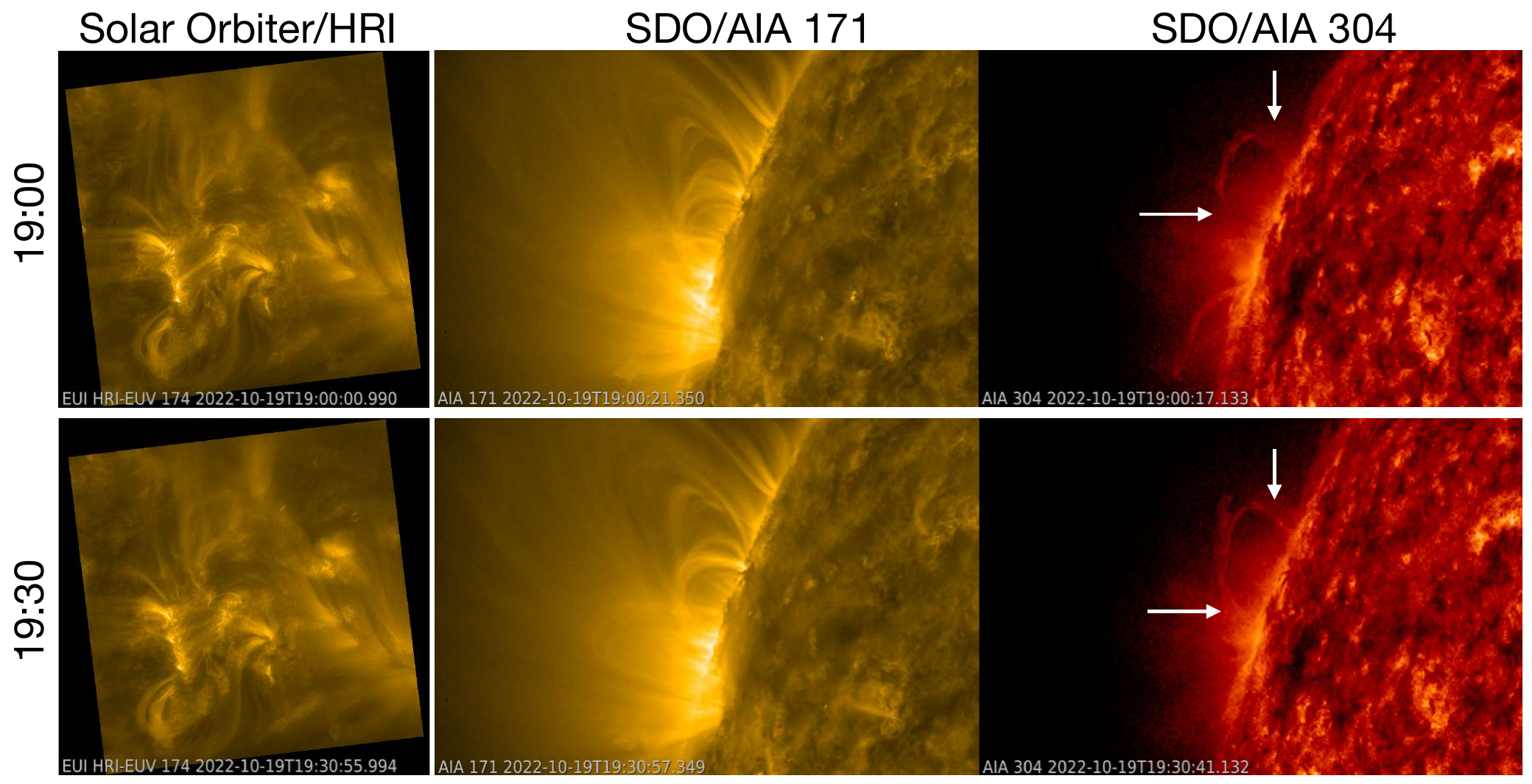} 
\caption{Coronal loops and coronal rain observed from two vantage points separated by 72\degree with Solar Orbiter/HRI and SDO/AIA. The upper row shows an observation obtained on 19 October 2022 at 19:00, and the bottom row shows an observation obtained 30 minutes later. Arrows highlight places where the coronal rains were observed.}
\label{fig:fig_coronal_rain}
\end{figure*}

\section{Observational highlights}
\label{sect:obs_highlights}

\subsection{Coronal loops}
\label{sect:obs_loop}

Coronal loops are the building blocks of the solar corona. They are rooted in regions with a strong magnetic field connecting opposite polarities and are composed of dense plasma compared to the coronal background. Analysing the detailed structure and evolution of coronal loops can provide crucial information on coronal heating mechanisms. Such review papers as \citet{Zaitsev2008} and \citet{Reale2014} discuss the properties and significance of coronal loops.
The coronal loops can interact with small-scale features. Thus, to answer question Q2, investigation of the coronal loops is also important.

Between 18 and 20 October 2022, we observed a coronal loop system at the solar limb in the \ion{Fe}{xiii} line with CryoNIRSP from Earth's perspective.  The system was located close to the disc centre (Fig.~\ref{fig:overview_fov_19oct}) from the Solar Orbiter perspective. 
The coronal intensity image shows an extended loop of a complex topology (Fig.~\ref{fig:overview_fov_19oct}).
These observations are good for studying the relation between the coronal magnetic field amplitude and the coronal rain frequency and amplitude.
The relationship between coronal loops and coronal rain implies that studying coronal loops is necessary to provide an answer to question Q3.

CryoNIRSP has opened a new possibility for investigating the characteristics of coronal loop properties, including the topology, dynamics, and density. The CryoNIRSP team is working to  provide tools for the coronal magnetic field analysis based on the \ion{Fe}{xiii} lines \citep{Schad2024b}.
The DKIST CryoNIRSP observations in coronal temperatures (\ion{Fe}{XIII}) show a loop system extended out of the solar limb that was part of the active region. The loop system grows in time between 18 and 19 October 2022. \citet{Petrova2024} investigated dynamics structures in this loop system using SPICE and HRI observations on 18 October 2022 and found a complex pattern in the Doppler velocity map. In this pattern, the blue- and red-shifted features, with a Doppler velocity of 20-200 km/s, are located opposite each other, which can suggest a rotating structure in the centre of the loop system. 

\citet[][private communication]{Mzerguat-submitted}  investigated the time evolution of the plasma composition in this region throughout the chromosphere all the way up to the corona by using Solar Orbiter/SPICE and Hinode/EIS data. They found different signatures of first ionisation potential bias enhancement depending on the location within the loop system by comparing short and closed loops compared with open-like loops. The complementary dataset from DKIST, especially on the magnetic field configuration and Doppler measurements with CryoNIRSP, could provide additional context to understand the origin of such enhancements, for example by characteristing the evolution of the rotating structure analysed in \citet{Petrova2024}.

The observed loop system provides a great potential for detailed analysis. The coordinated observations from multiple viewpoints allowed us to study the coronal loop evolution topology, morphology, and dynamics. Intensity maps from various instruments, such as EUI/FSI on board Solar Orbiter and SDO/AIA, as well as high-resolution spectroscopy techniques, provide detailed information on the structure, topology, plasma flow, and dynamics of coronal loops \citep[e.g.][]{Podladchikova2021}. The analysis methods of active region studies with stereoscopic and multi-instrument observations obtained with the Solar Orbiter and several space-based and ground-based observations are described in \citet{Janvier2023}. These methods can also be used to investigate the active region presented in this paper.

Magnetic field measurements obtained alongside stereoscopic views open the possibility of reconstructing the 3D topology of coronal loops, which has only been done on rare occasions \citep[see e.g.][]{Schad2016}. \citet{Judge2019} discussed the challenges and advancements in measuring solar surface magnetic fields without sign ambiguity. \citet{Valori2023} developed and successfully applied a method for the removal of the sign ambiguity in the azimuth using stereoscopic observations based on SO/PHI-HRT data for the first time.

On 21 and 24 October 2022, the loop system was located at the disc centre for DKIST and Solar Orbiter observers. DKIST provided ViSP and VBI observations.
Coordinated observations of DKIST, Solar Orbiter, and SDO are critical for addressing unresolved issues regarding the physics of coronal loops, especially those concerning the processes that drive their dynamics and movement, the relationship between coronal loops and small-scale structures, and the contribution of coronal loops to solar corona heating.

\subsection{Small-scale brightenings}

Small-scale brightenings (with spatial scales of less than 1~Mm) are commonly observed in the solar atmosphere from the photosphere to the corona (see, for example, \citealt{Young2018}). Indeed, one of the key early discoveries of Solar Orbiter was the detection of thousands of small-scale short-lived brightenings in the quiet-Sun in EUI/HRI images (see \citealt{Berghmans2021}). These events are similar in spatial and temporal scale to other EUV brightenings detected in plage regions using images obtained by Hi-C \citep{Peter2013, Barczynski2017}. The Solar Orbiter quiet-Sun EUV brightenings have been shown to have signatures in a range of transition region spectra (see e.g. \citealt{Nelson2023}). However, it is not known if such events are also detectable in active regions, and if they are, whether they have signatures in photospheric or chromospheric spectra is still unknown.

The coordinated Solar Orbiter and DKIST observations presented in this article will foster the crucial investigation of such topics. The EUI/HRI observation revealed evidence for the presence of numerous EUV brightenings within the active region on both 21 and 24 October 2022. On these dates, the VBI and ViSP imaging and spectro-polarimetric data could be studied to identify whether any photospheric or chromospheric response was present. Additionally, the PHI/HRT and ViSP magnetogram maps could be used for tomographic reconstruction by using the stereoscopic view of the magnetic field structure corresponding to brightenings.

The observation of small-scale brightenings such as those reported here will allow one of the common scientific objectives(Q2, see Sect.~\ref{sect:intro}) to be fulfilled, namely, identifying the role of small-scale features in active region evolution. The coordinated observations of Solar Orbiter and DKIST open up new possibilities for studying the physics of the EUV brightenings. The coordinated observations are especially important to understanding the main mechanisms responsible for the EUV brightenings in the solar corona, its role in coronal or transition region heating \citep{Dolliou2024} and energy balance, its relationship with magnetic field dynamics in the solar atmosphere, and the temporal changes in spatial distribution and physical properties of EUV brightening in active regions.

\subsection{Coronal rain}
\label{sect:coronal_rain}

Coronal rain is a non-equilibrium phenomenon, and it describes confined regions ('clumps') of cool ($10^{3} - 10^{5}$\,K), dense ($10^{10} - 10^{13}$\,cm$^{3}$), and partially ionised plasma in the solar atmosphere. In the optically thin corona, radiative losses increase with the square of the electron density. This means that areas with a higher electron density will cool more quickly than their surroundings. As a result, material will flow into these cooler regions, further increasing the density in a run-away process. As a result, clumps of relatively cool and dense plasma ('condensations') form and subsequently move towards the Sun's surface, guided by the magnetic field of coronal loops \citep{Muller2003, Muller2004}. 
In this phase, they can be observed in chromospheric and transition region lines \citep{DeGroof2004} as well as in 'coronal' passbands such as 174~\AA\ when high-density condensation regions absorb EUV radiation. The speed of rain clumps depends on the size and geometry of the coronal loop in which they originate as well as their formation location \citep{Schrijver2001, Muller2004}. Rain clumps typically have velocities of tens of kilometers per second and a broad velocity distribution between 10 to 200 km/s \citep{Schrijver2001, Antolin2012}, which can be explained by the counteracting effects of gravity and increased plasma pressure below the condensation region \citep{Muller2005}. The high spatial-temporal data from Solar Orbiter allows the substructure of the solar coronal rain to be investigated (e.g. \citet{Antolin2023})

Observations obtained on 19 October 2022 show coronal rain in the SDO/AIA 304 channel at the solar limb.
The same region was observed with Solar Orbiter high-resolution instruments from a 77$\degree$ angular separation relative to the Sun-Earth line.
EUI/HRI data open up the possibility of investigating the evolution of the plasma clumps for the first time with a temporal resolution of 5 sec and a high spatial resolution (2 pixels $= $238 km) with a stereoscopic view jointly using SDO/AIA.
 HMI and PHI allow for the study of the photospheric magnetic field topology of the structures related to the coronal rain and a stereoscopic view.
Figure~\ref{fig:fig_coronal_rain} shows the coronal rain observed with SDO/AIA 304, the co-spatial coronal loop system observed with SDO/AIA 171, and the corresponding observation obtained with EUI/HRI 174.
The coordinated observations of DKIST, Solar Orbiter, and SDO can help in the study of the origin of coronal rain and its properties, such as defined in Sect.~\ref{sect:intro}.

Several aspects of coronal rain are still the subject of scientific investigation, such as changes in plasma properties during coronal rain observations and the role of magnetic field topology and its evolution in the creation and evolution of coronal rain. Coordinated observations between DKIST and Solar Orbiter will be very helpful in addressing these topics as well.

\section{Summary and conclusions}\label{sect:conclusion}

In this article, we have provided an overview of the first coordinated observations obtained by Solar Orbiter and DKIST. Between 18 and 24 October 2022, numerous coordinated observations were executed (as well as several days of supporting uncoordinated observations) utilising the strengths of the state-of-the-art instruments hosted at both of these new facilities. The coordinated observations of Solar Orbiter and DKIST provide a unique opportunity to investigate the physics of the solar atmosphere, from the photosphere to the corona, with unprecedented spatial resolutions and a stereoscopic view. Both Solar Orbiter and DKIST data are publicly available (fully open-access), thereby helping foster collaboration within the entire solar physics community. 

To help summarise these data products, we have also provided a shortlist (though it is certainly not exhaustive by any means) of scientific topics that could be investigated using them. Specifically, we identified coronal loop physics, small-scale brightenings formation, and coronal rain dynamics as potential research topics. The coordinated high-resolution imaging, magnetic field measurements, and spectroscopy data, combined, will provide ideal datasets to tackle these exciting research areas. Not only this, but the high-resolution observations from different vantage points opens the possibility of tomographic reconstruction of coronal loops based on stereoscopic observations. By combining intensity maps (for example, CryoNIRSP maps from DKIST and EUI imaging from Solar Orbiter, when it was at quadrature with Earth), reconstruction of the topology of solar structuring (such as coronal loops) becomes possible. These observations are a good candidate for an initial study on the relation between coronal magnetic field amplitude and coronal rain frequency and amplitude. However, as the CryoNIRSP data represent very early DKIST data, a deeper investigation into the data calibration fidelity will be required. By combining magnetic field maps (e.g. from VISP at DKIST and PHI from Solar Orbiter), it is possible to provide multi-line spectropolarimetry, which allows one to study remote sensing data of the magnetic field at different altitudes in the photosphere. Moreover, the removal of the ambiguity using stereoscopic measurements(accounting for the $180^\circ$ ambiguity) is a separate possibility. Further combining these datasets with others collected by additional spacecraft (such as IRIS and Hinode) offers the potential for even greater benefits for the community.

Following the successful campaign from October 2022 described here, we conducted a second campaign of coordinated observations between DKIST and Solar Orbiter in October 2023 (which also returned fully open-access data). Due to bad weather conditions, the planned coordinated campaign in 2024 was unsuccessful. We are planning another coordinated observation campaign that should take place in 2025.

\begin{acknowledgements}
Solar Orbiter is a space mission of international collaboration between ESA and NASA, operated by ESA. We are grateful to the ESA SOC and MOC teams for their support. The EUI instrument was built by CSL, IAS, MPS, MSSL/UCL, PMOD/WRC, ROB, LCF/IO with funding from the Belgian Federal Science Policy Office (BELSPO/PRODEX PEA 4000134088, 4000112292, 4000136424, and 4000134474); the Centre National d’Etudes Spatiales (CNES); the UK Space Agency (UKSA); the Bundesministerium für Wirtschaft und Energie (BMWi) through the Deutsches Zentrum für Luft- und Raumfahrt (DLR); and the Swiss Space Office (SSO). 
The development of SPICE has been funded by ESA member states and ESA. It was built and is operated by a multi-national consortium of research institutes supported by their respective funding agencies: STFC RAL (UKSA, hardware lead), IAS (CNES, operations lead), GSFC (NASA), MPS (DLR), PMOD/WRC (Swiss Space Office), SwRI (NASA), UiO (Norwegian Space Agency).
The German contribution to SO/PHI is funded by the BMWi through DLR and by MPG central funds. The Spanish contribution is funded by AEI/MCIN/10.13039/501100011033/ and European Union “NextGenerationEU”/PRTR” (RTI2018-096886-C5, PID2021-125325OB-C5, PCI2022-135009-2, PCI2022-135029-2) and ERDF “A way of making Europe”; “Center of Excellence Severo Ochoa” awards to IAA-CSIC (SEV-2017-0709, CEX2021-001131-S); and a Ramón y Cajal fellowship awarded to DOS. The French contribution is funded by CNES. The research reported herein is based in part on data collected with the Daniel K. Inouye Solar Telescope (DKIST), a facility of the National Solar Observatory (NSO). NSO is managed by the Association of Universities for Research in Astronomy, Inc., and is funded by the National Science Foundation. Any opinions, findings and conclusions or recommendations expressed in this publication are those of the authors and do not necessarily reflect the views of the National Science Foundation or the Association of Universities for Research in Astronomy, Inc. DKIST is located on land of spiritual and cultural significance to Native Hawaiian people. The use of this important site to further scientific knowledge is done so with appreciation and respect.
\end{acknowledgements}

\bibliographystyle{aa}
\bibliography{aa}

\begin{appendix}
\onecolumn
\section{DKIST/ViSP observation}\label{sec:appendix_visp}

\begin{figure*}[h!]
\center{
\includegraphics[width=0.8\textwidth]{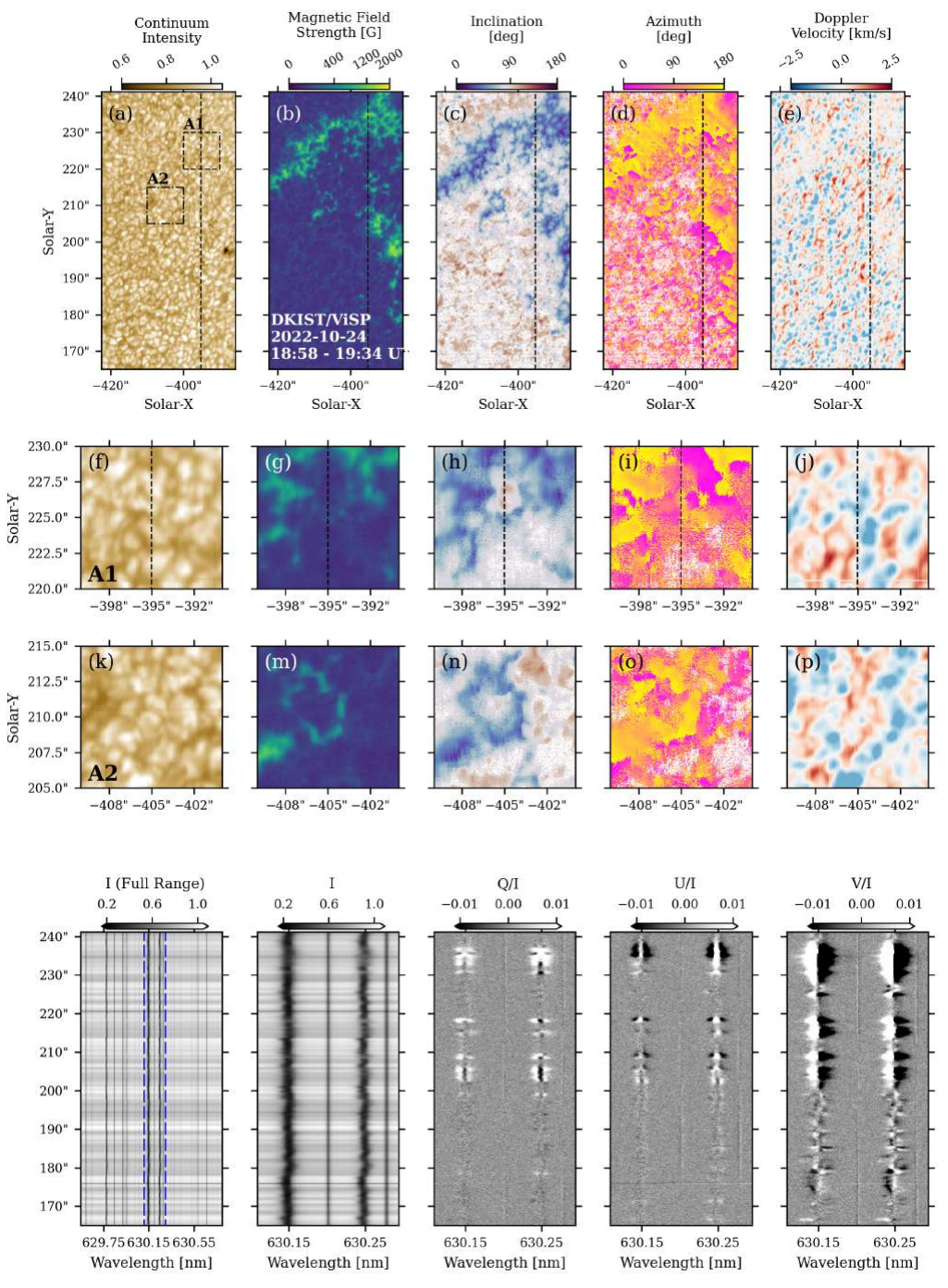} 
\caption{DKIST/ViSP spectropolarimetric data were acquired in the \ion{Fe}{i} 6300~\AA~bandpass on 24 October 2022 between 18:58 and 19:34 UT. The upper panels present results from Milne-Eddington inversions that yield the magnetic field vector. Panel (a) displays the photospheric granulation observed in continuum intensity. The corresponding vector magnetogram components highlight the magnetic field strength, which reaches values of up to 2 kG in panel (b), as well as the magnetic field inclination (c) and azimuth (d). The photospheric Doppler velocity is shown in panel (e), ranging from $\pm$~2.5 km/s. To further illustrate granular details, two selected regions, A1 and A2, are examined closely. The continuum intensity map, along with magnetic field strength, vector inclination, vector azimuth, and photospheric Doppler velocities for region A1 are shown in panels (f, g, h, i, j), while the corresponding data for region A2 are presented in panels (k, m, n, o, p). The Stokes vectors in the lower panel are extracted from the dashed vertical lines in the upper panels, providing additional insights into the data analysis.}
\label{fig:dkist_visp}}
\end{figure*}

The ViSP provides a detailed view of photospheric features. Figure~\ref{fig:dkist_visp} presents a zoom-in on two subregions within the ViSP FOV. The photospheric structures show nearly uniform granulation (Fig.~\ref{fig:dkist_visp}a). Additionally, the ViSP obtained vector spectropolarimetric data, enabling the construction of magnetic field strength maps (Fig.~\ref{fig:dkist_visp}b), which indicate magnetic field structures reaching up to 2 kG. The magnetic field inclination shown in Fig.~\ref{fig:dkist_visp}c is close to zero, while the magnetic field azimuth in Fig.~\ref{fig:dkist_visp}d approaches 180$^\circ$ in regions of peak magnetic field strength. The magnetic field azimuth is not disambiguated in these ViSP figures. Moreover, the photospheric Doppler velocity map in Fig.~\ref{fig:dkist_visp}e shows an almost unipolar structure, with velocities around $\pm$~2 km/s. Figures~\ref{fig:dkist_visp}f-j provide a closer look at region-of-interest (ROI) A1, presenting the same types of maps as seen in panels a-e. Similarly, panels k-p zoom in on ROI A2, displaying corresponding maps to those in panels a-e. Both regions A1 and A2 offer detailed insights into granule structures, their associated magnetic fields, and plasma flow properties. In addition, DKIST/ViSP provides information about the Stokes parameters. We focus here on the Stokes parameters along the dashed line at the bottom of Fig.~\ref{fig:dkist_visp}; the individual panels illustrate zoomed views of the I, Q/I, U/I, and V/I Stokes parameters.

\section{Summary of Solar Orbiter and DKIST datasets}

\begin{table*}

\caption{Summary of the successfully acquired Solar Orbiter datasets coordinated with DKIST. } 

\label{table:solOrb_obs}

\centering 
\begin{tabular}{p{1.5cm} c c c c c p{2.2cm}} 

\hline\hline 
{Instrument} & {Wavelength channel(s)} & {Start-End [UT]} & {Pointing [\arcsec]} & {1 px size [km]} & {FOV [Mm]} & {Cadence[s]}\\
\hline\hline 
\multicolumn{2}{l}{\textit{18 October 2022}} \\
EUI/FSI & 174, 304 & 00:00 - 23:50 & (-1187\arcsec, 916\arcsec) & 1040 & 3161$\times$3194 & 10 min; 20 min\\
SPICE & SW, LW & 07:55 - 23:37 & (-1256\arcsec, 748\arcsec) & x:1405, y:257 & 180$\times$214 & 11 min 29 sec \\
PHI/HRT & - & 03:15 - 23:15 & (-1265\arcsec, 791\arcsec) & 117 & 240$\times$240 & 1h \\
\hline 
\multicolumn{2}{l}{\textit{19 October 2022}} \\
EUI/FSI & 174, 304, & 00:00 - 23:50 & (-812\arcsec, 826\arcsec) & 1071 & 3257$\times$3291 & 10 min, 20 min \\
EUI/HRI & 174, Ly-$\alpha$ & 19:05 - 20:05 & (-1060\arcsec, 784\arcsec) & 119 & 243$\times$243 & 5 s., 30 s.\\
SPICE & SW, LW & 07:40 - 22:36 & (-881\arcsec, 661\arcsec) & x:1448,y:265& 185$\times$220 & 11 min 29 sec\\
PHI/HRT &-& 00:15 - 23:15 & (-889\arcsec, 704\arcsec) & 121 &247$\times$247 & 1h \\
\hline
\multicolumn{2}{l}{\textit{20 October 2022}} \\
EUI/FSI & 174, 304 & 00:00 - 23:50 & (-388\arcsec, 789\arcsec) & 1106 & 3363x3398 & 10 min, 20 min\\
EUI/HRI & 174, Ly-$\alpha$ & 19:05 - 20:05 & (-637\arcsec, 750\arcsec) & 123 & 251$\times$251 & 5 s., 30 s.\\
SPICE & SW, LW & 08:00 - 22:44 & (-456\arcsec, 620\arcsec) & x:1495, y:274 & 191$\times$228 & 11 min 29 sec\\
PHI/HRT &-& 03:15 - 23:15 & (-464\arcsec, 663\arcsec) & 125& 256$\times$256 & 1h \\
\hline 
\multicolumn{2}{l}{\textit{21 October 2022}} \\
EUI/FSI & 174, 304 & 00:00 - 23:50 & (21\arcsec, 730\arcsec) & 1143 & 3476$\times$3512 & 10 min, 20 min\\
EUI/HRI & 174, Ly-$\alpha$ & 19:05 - 20:05 & (-228\arcsec, 691\arcsec) & 127 & 259$\times$259 & 5~s., 30~s.\\
SPICE & SW, LW & 07:55 - 22:39 & (-46\arcsec, 567\arcsec) & x:1545, y:283 & 198$\times$235 & 11 min 29 sec\\
PHI/HRT &-& 00:15 - 23:15 & (-57\arcsec, 610\arcsec) & 129 & 264$\times$264 &1h\\
\hline 
\multicolumn{2}{l}{\textit{24 October 2022}} \\
EUI/FSI & 174, 304 & 00:00 - 23:50 & (1134\arcsec, 641\arcsec) & 1265 & 3846$\times$3886 & 10 min, 20 min \\
EUI/HRI & 174, - & 19:05 - 19:35 & (884\arcsec, 609\arcsec) & 140 & 287$\times$287 & 5 s\\
SPICE & SW, LW & 07:55 - 21:53 & (1070\arcsec, 476\arcsec) & x:1710, y:313 & 219$\times$260 & 11 min 29 sec\\
PHI/HRT &-& 03:15 - 23:20 & (1069\arcsec, 521\arcsec) & 143 & 219$\times$219 & 1h\\
\hline 

\end{tabular}
\tablefoot{The target coordinates correspond to the centre of the FOV of the Solar Orbiter instruments in helioprojective coordinates as viewed from Solar Orbiter. The distance between the Sun and Solar Orbiter on each day are defined in the Table~\ref{table:obs_setup}. The table presents UT measured at the Earth. Pointing information are presented in the table at 19:05 UT for each coordinated observation day (i.e. when the high-resolution coordinated observations began). SPICE provided two types of observations. The table presents details for \texttt{SCI\_DYN-MED\_SC\_SL04\_5.0S\_FF} SPICE observations. The SPICE \texttt{SCI\_DYN-MED\_SC\_SL04\_5.0S\_FF} observations were provided with duration of $\approx$3.25h, but out of DKIST observation time range (thus we do not show its details in the table). The pointing position in arcsec depends on the distance of the spacecraft to the Sun, and distance of the Solar Orbiter changed between 0.32 and 0.39 au for observations presented in this table.}
\end{table*}

\begin{table*}

\caption{Summary of acquired DKIST spectroscopic or spectropolarimetric datasets.} 

\label{table:dkist_obs} 

\small
\centering 

\begin{tabular}{llccclc} 
\hline\hline 
{Instrument} &{Wavelength [\AA]} & {Start-End [UT]} & {DKIST Pointing [\arcsec]} & {FOV [Mm]} & {Mode} & {Raster Step[\arcsec]}\\ 

\hline\hline 
\multicolumn{2}{l}{\textit{18 October 2022}} \\
\multirow{6}{*}{CryoNIRSP} & \ion{Fe}{xiii} 10798 & 18:30-18:42 & \multirow{6}{*}{(-$1023\arcsec$,$293\arcsec$)} & \multirow{6}{*}{145x167} & Spectroscopy(OP1) & 0.5 \\
& \ion{Fe}{xiii} 10746 & 19:03-19:56 & & & Polarimetry (OP2) & 2 \\
& \ion{Fe}{xiii} 10798 & 19:57-20:10 & & & Spectroscopy (OP1) & 0.5 \\
& \ion{Fe}{xiii} 10746 & 20:11-20:23 & & & Spectroscopy (OP1) & 0.5 \\
& \ion{Fe}{xiii} 10746 & 20:24-20:46 & & & Deep-polarimetry(OP3) & 20 \\
& \ion{Fe}{xiii} 10746 & 21:10-21-22 & & & Spectroscopy (OP1) & 0.5 \\

\hline 
\multicolumn{2}{l}{\textit{19 October 2022}} \\
\multirow{9}{*}{CryoNIRSP} & \ion{Fe}{xiii} 10798 & 19:09-19:22 & \multirow{9}{*}{(-$1023\arcsec$,$293\arcsec$)} & \multirow{9}{*}{145x167} & Spectroscopy (OP1) & 0.5 \\
& \ion{Fe}{xiii} 10746 & 19:22-19:34 & & & Spectroscopy (OP1) & 0.5 \\
& \ion{Fe}{xiii}10746 & 19:52-20:45 & & & Polarimetry (OP2) & 2 \\
& \ion{Fe}{xiii} 10746 & 21:02-21:25 & & & Deep-polarimetry (OP3) & 20 \\
& \ion{Fe}{xiii} 10798 & 21:27-21:40 & & & Spectroscopy (OP1) & 0.5 \\
& \ion{Fe}{xiii} 10746 & 21:41-21:52 & & & Spectroscopy (OP1) & 0.5 \\
& \ion{Fe}{xiii} 10746 & 22:04-22:57 & & & Polarimetry (OP2) & 2 \\
& \ion{Fe}{xiii} 10798 & 23:04-23:16 & & & Spectroscopy (OP1) & 0.5 \\
& \ion{Fe}{xiii} 10746 & 23:17-23:29 & & & Spectroscopy (OP1) & 0.5\\

\hline
\multicolumn{2}{l}{\textit{20 October 2022}} \\
\multirow{5}{*}{CryoNIRSP} & \ion{Fe}{xiii} 10798 & 18:41-18:53 & \multirow{5}{*}{(-$1023\arcsec$,$293\arcsec$)} & \multirow{5}{*}{145x167} & Spectroscopy (OP1) & 0.5 \\
& \ion{Fe}{xiii} 10746 & 18:54-19:06 & & & Spectroscopy (OP1) & 0.5\\
& \ion{Fe}{xiii} 10746 & 19:20-20:14 & & & Polarimetry (OP2) & 2 \\
& \ion{Fe}{xiii} 10746 & 20:47-21:10 & & & Deep-polarimetry (OP3) & 20 \\
& \ion{Fe}{xiii} 10798 & 21:25-21:36 & & & Spectroscopy (OP1) & 0.5 \\

\hline 
\multicolumn{2}{l}{\textit{21 October 2022}} \\
\multirow{2}{*}{ViSP} & \multirow{2}{*}{Fe I 6300} & 18:44-19:37 & (-$834\arcsec$,$230\arcsec$) & 39x54 & \multirow{2}{*}{Polarimetry} & \multirow{2}{*}{0.0536}\\
& & 20:24-21:17 & (-$841\arcsec$,$214\arcsec$) & 39x54 & & \\

\hline 
\multicolumn{2}{l}{\textit{24 October 2022}} \\
\multirow{4}{*}{ViSP} & \multirow{4}{*}{Fe I 6300} & 18:28-18:54 & (-$397\arcsec$,$198\arcsec$) & 19x54 & \multirow{4}{*}{Polarimetry} & \multirow{4}{*}{0.0536} \\
& & 18:57-19:34 & (-$397\arcsec$,$198\arcsec$) & 28x54 & & \\
& & 19:47-20:39 & (-$397\arcsec$,$198\arcsec$) & 27x54 & & \\
& & 21:28-22:20 & (-$379\arcsec$,$185\arcsec$) & 39x54 & & \\

\hline 

\end{tabular}
\tablefoot{Pointing or target coordinates correspond to the centre of the DKIST FOV in helioprojective coordinates (X,Y). Imaging data from the CryoNIRSP context imager (He I 10830~\AA) and from VBI (Blue and Red) were acquired in parallel with the CryoNIRSP spectrograph and VISP, respectively, as described in the text. }
\end{table*}

\end{appendix}
\end{document}